\documentclass{aa}
\usepackage{txfonts}
\usepackage[dvips]{graphicx}
\usepackage{natbib}
\bibpunct{(}{)}{;}{a}{}{,}

\titlerunning{Galaxy evolution by color-log($n$) type since $z \sim 1$ in the UDF}

\title{Galaxy evolution by color-log($n$) type since redshift unity in the Hubble Ultra Deep Field}

\author{E. Cameron \and S. P. Driver}

\institute{SUPA\thanks{Scottish Universities Physics Alliance}, School of Physics and Astronomy, University of St Andrews, North Haugh, St Andrews, KY16 9SS, Scotland}

\date{Received xxx / Accepted 29 October 2008}

\abstract{}{We explore the use of the color-log($n$) (where $n$ is the global S\'ersic index) plane as a tool for subdividing the galaxy population in a physically-motivated manner out to redshift unity.  We thereby aim to quantify surface brightness evolution by color-log($n$) type, accounting separately for the specific selection and measurement biases against each.}{We construct ($u$-$r$) color-log($n$) diagrams for distant galaxies in the Hubble Ultra Deep Field (UDF) within a series of volume-limited samples to $z=1.5$.  The color-log($n$) distributions of these high redshift galaxies are compared against that measured for nearby galaxies in the Millennium Galaxy Catalogue (MGC), as well as to the results of visual morphological classification.  Based on this analysis we divide our sample into three color-structure classes.  Namely, `red, compact', `blue, diffuse' and `blue, compact'.  Luminosity-size diagrams are constructed for members of the two largest classes (`red, compact' and `blue, diffuse'), both in the UDF and the MGC.  Artificial galaxy simulations (for systems with exponential and de Vaucouleurs profile shapes alternately) are used to identify `bias-free' regions of the luminosity-size plane in which galaxies are detected with high completeness, and their fluxes and sizes recovered with minimal surface brightness-dependent biases.  Galaxy evolution is quantified via comparison of the low and high redshift luminosity-size relations within these `bias-free' regions.}{We confirm the correlation between color-log($n$) plane position and visual morphological type observed locally and in other high redshift studies in the color and/or structure domain.  The combined effects of observational uncertainties, the morphological K-correction and cosmic variance preclude a robust statistical comparison of the shape of the MGC and UDF color-log($n$) distributions.  However, in the interval 0.75$<$$z$$<$1.0 where the UDF $i$-band samples close to rest-frame $B$-band light (i.e., the morphological K-correction between our samples is negligible) we are able to present tentative evidence of bimodality, albiet for a very small sample size (17 galaxies).  Our unique approach to quantifying selection and measurement biases in the luminosity-size plane highlights the need to consider errors in the recovery of both magnitudes and sizes, and their dependence on profile shape.  Motivated by these results we divide our sample into the three color-structure classes mentioned above and quantify luminosity-size evolution by galaxy type.  Specifically, we detect decreases in $B$-band, surface brightness of $1.57 \pm 0.22$ mag arcsec$^{-2}$ and  $1.65 \pm 0.22$ mag arcsec$^{-2}$ for our `blue, diffuse' and `red, compact' classes respectively between redshift unity and the present day.}{}


\begin{document}
\maketitle

\section{Introduction}
Recent results from wide field imaging surveys, such as the Sloan Digital Sky Survey (SDSS; \citealt{ade06}) and the Millennium Galaxy Catalogue (MGC; \citealt{lis03}), have highlighted the bimodal nature of the local galaxy population.  In particular, bright galaxies at $z \sim 0$ naturally divide into two distinct types in the color-magnitude \citep{str01}, log($n$)-magnitude \citep{dri06,bal06} and color-log($n$) \citep{dri06} planes\footnote{Here $n$ refers to the global `S\'ersic index', i.e., the inverse-exponent in the best-fit S\'ersic model to the galaxy light distribution.  The S\'ersic model is ``most commonly expressed as an intensity profile, such that $I(R) = I_e \exp \left\{ -b_n \left[ \left( \frac{R}{R_e} \right)^{1/n} - 1 \right] \right\}$, where $I_e$ is the intensity at the effective radius $R_e$ that encloses half of the total light from the model'' \citep{gra05} and $b_n$ is a normalization constant.}.  One type displays red colors consistent with old stellar populations (and/or very high dust contents) and the compact light profiles typical of pressure-supported systems.  The other type has bluer colors indicative of more recent star-formation and younger stellar populations, and the diffuse light profiles of rotationally-supported systems.  \citet{dri06} demonstrate that the red, compact peak generally corresponds to morphologically early-type systems (E/S0s), while the blue, diffuse peak is dominated by late-type systems (Sd/Irr), with intermediate spirals (Sa-Sc) spanning both. This bimodality suggests that there are two distinct pathways of galaxy evolution and formation.  It is incumbent upon theories of galaxy formation to explain the origin and nature of these two pathways, and the evolution of galaxies within and between them.  In the orthodox hierarchical clustering (HC) scenario the blue sequence (late-type population) is generally thought to form via the cooling of gas in rotating dark matter halos spun up by tidal torques in the early universe \citep{pee69,fal80}.  The red sequence (early-type population) is then built up via mergers between blue sequence members \citep{too72,too77} and dry mergers amongst the red sequence \citep{van99,kho03}. Fading of outer discs in spirals to reveal a central bulge, or conversely enhancement of the bulge relative to the disc, may also be an important evolutionary pathway (\citealt{chr04} and references therein).  Secular evolution, such as the formation of bars and bulges via disc instabilities (see \citealt{kor04} for a recent review) and perhaps the formation of outer disc truncations \citep{deb06}, will likely play a significant role in isolated systems on long timescales.\\

Deep imaging surveys, such as the Hubble Ultra Deep Field (UDF), allow us to constrain quantitative predictions from galaxy formation theories and gauge the importance of specific evolutionary mechanisms at different epochs.  For instance, \citet{mci05} compare high redshift galaxies in the GEMS survey to those at low redshift in the SDSS and find that the luminosity-size and mass-size relations are consistent with passive ageing of old stellar populations since $z \sim 1$.  By a further comparison to theoretical models they limit the role of dissipationless merging to $<$1 major merger over this timescale, favoring instead the role of disc fading in order to build up the red sequence number density.  Alternatively, \citet{bla06} examines evolution of the luminosity functions of the red and blue sequences from $z \sim 1$ in DEEP2 to $z \sim 0$ in the SDSS and finds negligible change in the number density on the blue sequence (to within 10\%).  He argues that this supports the need for dry mergers to evolve the red sequence number density given the lack of luminous blue progenitors. The crucial elements of this genre of study are the system adopted for assigning galaxy type and the methodology employed to assess selection and measurement biases.  Recent work has favored physically-motivated, objective and quantitative classification systems for the former.  For instance, one can make use of the correlation between morphological type and S\'ersic index \citep{tru06} or color and S\'ersic index \citep{mci05} to separate early-type and late-type galaxies.  In order to maximize the reliability of the division between physical types one might attempt to identify each peak of the bimodality and place the division across the corresponding saddle point at each epoch.  This approach was successfully applied by \citet{bel04} and \citet{bla06} studying the high redshift color-magnitude bimodality, but has yet to be applied to the joint color-log($n$) bimodality.\\

Understanding the nature of the selection effects that operate on any given imaging survey is crucial to the robust interpretation of the data.  These include biases against the detection and photometry of compact, extended, faint and/or low surface brightness systems.  This is well-documented in the literature (e.g.\ \citealt{dis76,dis83,imp97,dri99,cro02,dri05}).  The low surface brightness limit is of vital importance to studies that attempt to recover the properties of apparently faint and/or dim galaxies, whether at the faint end of the luminosity function at low redshift or for intrinsically bright galaxies at high redshift.  As \citet{dal98} and \citet{cro02} point out, low surface brightness systems suffer from both a Malmquist incompleteness bias and from an under-estimation of their total fluxes---and the dual impact of these sources of error can conspire to mimic the effect of galaxy evolution with redshift.  For instance, \citet{sim99} and \citet{rav04} argue that the surface brightness evolution reported in the first generation of Hubble Space Telescope (HST) imaging survey studies (e.g. \citealt{sch95,lil98,roc98}) was almost entirely the result of selection bias.  More recent, deep, HST, imaging studies (such as \citealt{mci05} and \citealt{tru06}) have adopted improved approaches to handling selection bias (and do, in fact, find evidence for a significant degree of surface brightness evolution to $z \sim 1$).  However, best practice methods are not applied universally and there is as yet no clear consensus on the best way to evaluate surface brightness bias.  In \citet{cam07} we introduced a relatively conservative approach for identifying bias-free regions in the luminosity-size plane.  We demonstrated that even in the exceptionally deep UDF imaging the systematic under-estimation of total flux by Kron-like aperture magnitudes at low surface brightness could be causing bright, yet diffuse, systems to be recovered as faint, compact objects.  However, in that study we limited our investigation to the case of exponential, disc-like profiles, but for more concentrated galaxies the biases are known to be much worse \citep{gra05}.  It is, therefore, important to extend this analysis to `elliptical-like', de Vaucouleurs profile shapes in order to properly quantify evolution in the `red, compact' population.\\

In this paper we construct the color-log($n$) distribution of galaxies out to $z \sim 1.5$ in the Hubble Ultra Deep Field (UDF) and compare it against that of the local galaxy population in the Millennium Galaxy Catalogue (MGC).  We examine the use of this parameter space for morphological classification and search for evidence of bimodality.  Dividing our samples into `blue, compact', `blue, diffuse' and `red, compact' classes, we then investigate evolution in luminosity-size relations of the latter two (predominant) classes.  Careful attention is paid to the relevant selection effects and uncertainties, which we evaluate via artificial galaxy simulations.  The outline of the paper is as follows: in Section 2 we describe the construction of our dataset; in Section 3 we quantify selection effects and measurement biases; in Section 4 we present our results and discuss the relevant uncertainties; in Section 5 we compare our findings to other recent studies and discuss their interpretation in light of the hierarchical formation paradigm; and in Section 6 we summarise our conclusions.  A cosmological model with $\Omega_{M} = 0.3$, $\Omega_{\Lambda} = 0.7$ and $H_{0} = 100$ km s$^{-1}$ Mpc$^{-1}$ is used throughout.  These specific values of the cosmological parameters were adopted for ease of comparison between the present UDF work and the slightly older MGC results.  Unless otherwise stated, all magnitudes are given in the AB system.\\

\section{Data}
The Hubble Ultra Deep Field \citep{bec06} consists of an 11 arcmin$^2$ patch of sky centred on RA = $03\degr$ $32'$ $39.0''$, Dec = $-27\degr$ $47'$ $29.1''$ (J2000) in the region of the Fornax Constellation.  The publicly released ACS/WFC Combined Images (Version 1.0) span the optical wavelength range 3700 to 10,000$\AA$ in four wide-band filters: F435W ($B$), F606W ($V$), F775W ($i$) and F850LP ($z$).  The F775W $i$-band image has the longest total exposure time of 347,110 s (144 orbits).  Each single exposure was half an orbit in duration with the pointing cycled through a four part dither pattern.  Each image has been processed through the standard HST data pipeline and drizzled to a pixel scale of 0.03$''$ pixel$^{-1}$.  The PSF FWHM is $\sim$0.$''$081 in the $i$-band image \citep{bec06}, which is close to the theoretical diffraction limit of the HST at these wavelengths.  Supplementary observations in the near-IR have been made with NICMOS in the F110W ($J$) and F160W ($H$) filters \citep{tho05}, although only for a reduced area of $\sim$5.76 arcmin$^2$ within the UDF region.  These images are drizzled to a coarser pixel scale (0.$''$09 pixel$^{-1}$) appropriate to the larger PSF of the NIC 3 camera ($\sim$0.$''$36 FWHM in the $J$-band).\\

\subsection{The iUDF-BRIGHT Sample, Kron Magnitudes and Half Light Radii}
In \citet{cam07} we presented iUDF-BRIGHT---a catalogue of 2497 sources brighter than 28th magnitude in the ACS $i$-band image.  These were identified using \textsc{SExtractor} \citep{ber96} with a constant surface brightness threshold of 27.395 mag arcsec$^{-2}$ and were matched, where possible, with counterparts in the Version 1.0 data release (B04\footnote{In \citet{cam07} we called this `the on-line catalogue', but here we adopt the syntax of \citet{coe06} for simplicity.  Hence, we also now refer to \citet{tho05}'s $J$$+$$H$ catalogue as `T05'.}) catalogue to compare object magnitudes and deblending.  Our isophotal magnitudes were found to be in good agreement with those in B04 with the largest mean difference just $0.06 \pm 0.07$ mag in the worst 27-28th magnitude bin.  A total of 125 objects were found to have conflicting centroid positions, so these were visually inspected.  The B04 deblending was preferred in 60 instances, our deblending preferred in 56 instances and there were 9 spurious detections.  Elliptical aperture, major axis, half light radii were then computed for the entire iUDF-BRIGHT sample using Kron magnitude fluxes and a crude PSF correction.\\

\subsection{Colors, Photometric Redshifts and S\'ersic Indices}
\citet{coe06} present magnitudes, photometric redshifts and S\'ersic indices for a large sample of objects in the ACS and NICMOS UDF images.  Their catalogue (C06) includes both the B04 $i$ and $z$-band sources and the T05 $J$$+$$H$-band sources, as well as their own new $B$$+$$V$$+$$i$$+$$z$ and $J$$+$$H$ detections.  The magnitudes provided by \citet{coe06} for each filter are well suited for computing galaxy colors as they are aperture-matched, PSF-corrected, and corrected for the difference between isophotal and Kron fluxes.  Photometric redshifts are estimated by \citet{coe06} using the Bayesian BPZ code \citep{ben00}.  They adopt the SED library of \citet{ben04} with the addition of two new templates representing young, blue, star-forming galaxies.  Their photometric redshifts agree well with those determined spectroscopically for 41 bright galaxies in the UDF with an RMS of $\Delta z = 0.04 (1 + z_{\mathrm{spec}})$ upon exclusion of 4 outliers.  However, as we demonstrated in \citet{cam07} by comparison to an alternative UDF photometric redshift catalogue (M05, supplied by B.\ Mobasher, priv.\ comm.), the redshift estimates derived using this technique are highly sensitive to the choice of SED library and input magnitude type, particularly for faint galaxies.  This contributed a large source of uncertainty to our previous results, although we restrict our investigation to bright galaxies with $M_B < -18$ mag and $0.25 < z < 1.5$ for which the redshifts should be the most reliable.  In fact, 80\% of galaxies within these limits have their C06 and M05 photometric redshift estimates in agreement within $\pm$0.1$(1+z_{\mathrm{av}})$.  Finally, \citet{coe06} provide S\'ersic indices for all objects in the B04 catalogue.  These are global S\'ersic indices computed via 2D model fits to each galaxy's $i$-band image using the \textsc{galfit} software package.\\

As mentioned above, in \citet{cam07} we matched objects in the iUDF-BRIGHT sample to those in B04 to check our source extraction and photometry.  Since \citet{coe06} preserve B04's ID numbering system, it was a simple process to link the C06 morphological and photometric catalogues to our galaxy catalogue.  None of the 56 galaxies for which we preferred an alternative deblending to that of B04 have redshifts in C06 that place them within the limits of this study.  Hence, their lack of a proper match does not affect our work.\\

\subsection{K-Corrections, Absolute Magnitudes and Rest-Frame Colors}
Individual galaxy K-corrections are needed to compute rest-frame absolute magnitudes and colors.  Here we follow a similar procedure to \citet{cam07}, fitting redshifted spectral templates from the library of \citet{pog97} to each galaxy's observed optical and near-IR (where available) broadband photometry.  However, this time we use the aperture-matched, PSF-corrected fluxes and \textsc{BPZ} redshifts from the C06 catalogue.  The magnitudes used in our luminosity-size relations are all derived by K-correcting from the observed $i$-band (to rest-frame MGC $B$-band) because we have performed detailed modelling of the detection completeness and photometric reliability of galaxies in the ACS $i$-band image.  However, when deriving rest-frame $(u-r)$ colors we utilize bandpass shifting, i.e., K-correcting from the observed filters sampling nearest to rest-frame $u$ and $r$-band light at each redshift (either side of the 4000$\AA$ break).  Hence, we correct to $(u-r)_{\mathrm{rest}}$ from $(B-i)_{\mathrm{obs}}$ at $0.25 < z < 0.5$, $(V-z)_{\mathrm{obs}}$ at $0.5 < z < 1.0$ and $(i-J)_{\mathrm{obs}}$ (if available, otherwise $(V-z)_{\mathrm{obs}}$) at $1.0 < z < 1.5$.  Absolute magnitudes and colors are then derived using these K-corrections and a cosmological model with $\Omega_{M} = 0.3$, $\Omega_{\Lambda} = 0.7$ and $H_{0} = 100$ km s$^{-1}$ Mpc$^{-1}$.  These specific values of the cosmological parameters were adopted in order to facilitate comparison to the earlier, published MGC data.\\

\section{Selection Effects}
Here we quantify the selection effects and systematic measurement biases for galaxies in our sample in the luminosity-size plane.  Our approach is more conservative than other recent studies (e.g. \citealt{bou04,sar07}) because we use artificial galaxy simulations to test not only object completeness, but size and magnitude recovery as well.  We then select only the region of the luminosity-size plane at each redshift where the detection completeness of our simulated galaxies is over 90\% and at least 75\% of these have measured fluxes and half light radii within 10\% of their input (i.e., intrinsic) values.  We previously applied this methodology successfully to our study of the $z \sim 0.75$ UDF galaxy population (not subdivided by galaxy type) in \citet{cam07}, but only using exponential ($n=1$) profile shapes (on the assumption that our sample was dominated by disc-dominated galaxies at this redshift).  Here we extend this work to cover compact, bulge-dominated galaxies by additionally performing simulations using de Vaucouleurs ($n=4$) profile shapes, and contrasting the results against those from our exponential profile simulations.\\

For our UDF galaxy photometry we use \textsc{SExtractor}'s Kron magnitudes and half light radii.  Kron magnitudes \citep{kro80} employ a luminosity-weighted average to define the so-called Kron radius ($R_1$), and the corresponding magnitude is computed by summing the flux within an aperture of radius $2.5 R_1$.  Theoretically, the Kron magnitude should contain 96.0\% of the flux in an exponential profile and 90.4\% of the flux in a de Vaucouleurs profile (integrated to infinity).  However, in practice these theoretical values are rarely obtained because of observational constraints.  As \citet{gra05} point out `what is important is not the sky-level or isophotal-level one reaches, but the number of effective radii that have been sampled'.  It is this source of error that we are effectively trying to gauge via our artificial galaxy simulations.  Of course, when we refer to our ability to accurately recover the total or true flux of a de Vaucouleurs type profile we mean the total \textit{Kron} flux and are allowing for the 9.6\% loss against the integration of the profile to infinity.  As Kron magnitudes are used in both the UDF and MGC photometry this gap between the total integrated and total Kron magnitudes will not introduce a source of error into comparisons of their relative magnitude, size and surface brightness distributions.  It is also important to note that errors in the measured Kron magnitude will flow on to cause errors in the measured half light radius, with an under-estimation of flux leading to an under-estimation of size.  The extent of this under-estimation again varies with profile shape. Hence, we test our ability to accurately recover both magnitudes and sizes here.\\

\subsection{Artificial Galaxy Simulations and Apparent Selection Limits}
Our procedure for performing these selection bias tests is outlined in detail in \citet{cam07}.  Essentially, it involves dividing the observed luminosity-size plane into a grid and at each grid point generating 100 artificial galaxies with random inclinations and axial ratios using the \textsc{IRAF artdata} package.  These are inserted into the UDF $i$-band image, and \textsc{SExtractor} used to search for them and to measure Kron magnitudes.  Half light radii are then computed using the elliptical aperture method.  The resulting detection positions, magnitudes and sizes are compared to those in the input object list.\\

\begin{figure}
\center
\includegraphics[width=7.01cm]{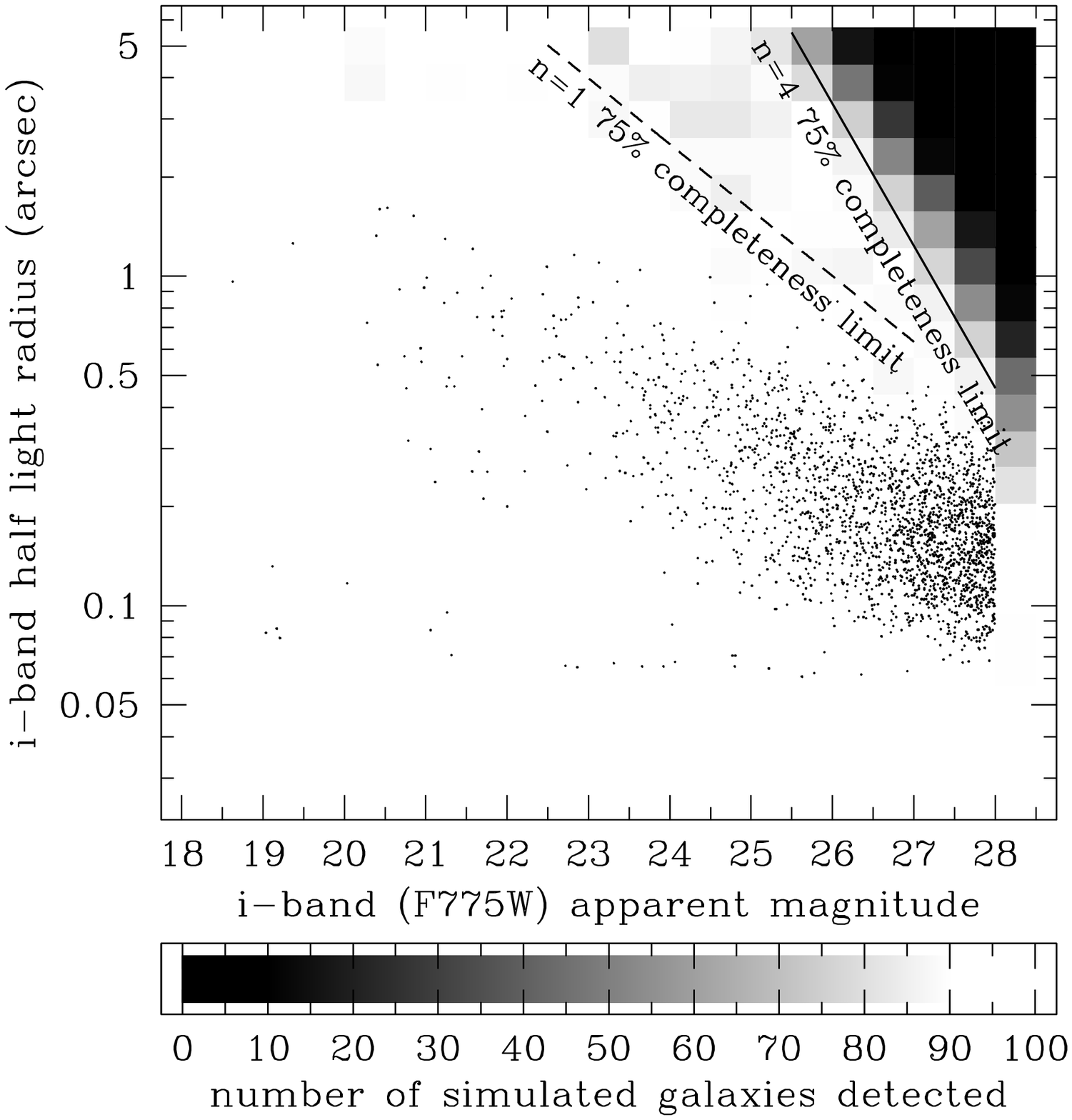}
\includegraphics[width=7.01cm]{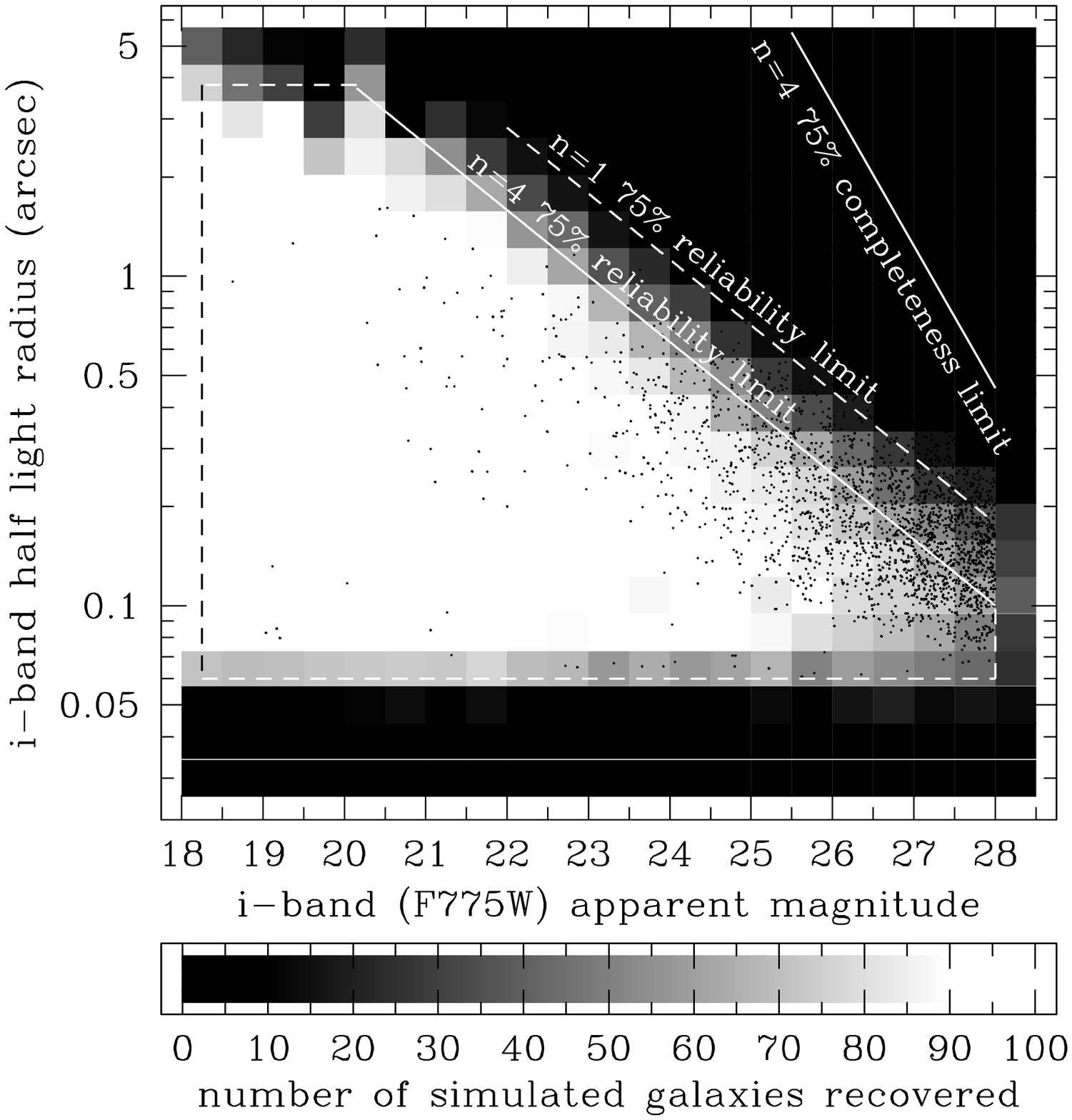}
\includegraphics[width=7.01cm]{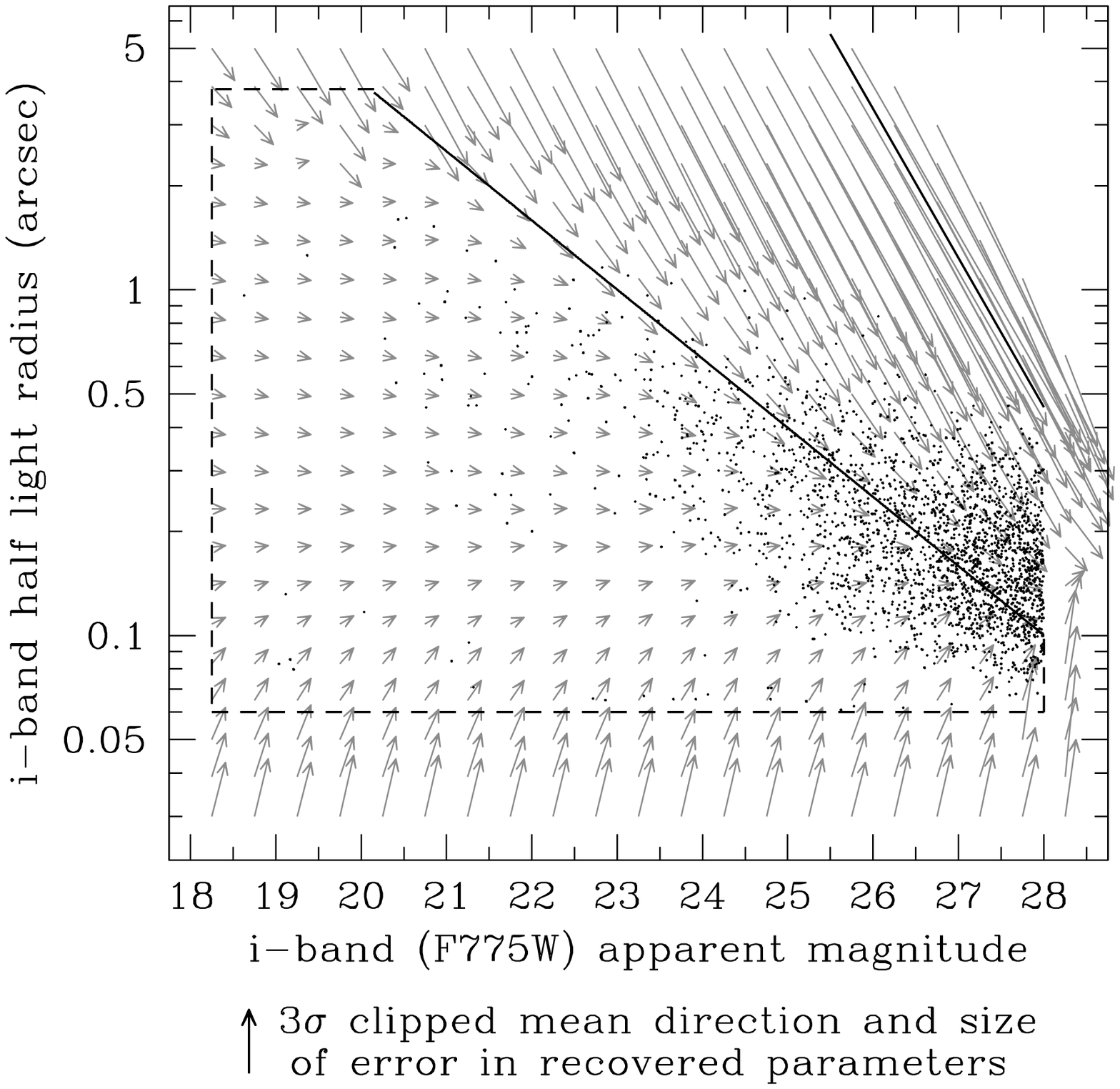}
\caption{\label{dev} Completeness, `reliability' and `error vector' diagrams for de Vaucouleurs ($n=4$) profile objects in the ACS UDF $i$-band image computed from artificial galaxy simulations as described in Section 3.1.  The greyscale squares and the arrows indicate the simulation results and the black dots show the distribution of the real observed data.  The 75\% completeness and reliability limits that define our bias-free region are marked for clarity.  The equivalent limits for exponential ($n=1$) type galaxies at low surface brightness are also indicated for comparison.}
\end{figure}

Figure \ref{dev} displays the results of our simulations for galaxies with de Vaucouleurs ($n=4$) profile shapes in a set of three plots.  The first plot indicates our detection completeness as a function of apparent magnitude and effective radius (i.e., the number of artificial galaxies detected of the 100 inserted at each grid point).  The second plot reveals our `reliability' in this parameter space, which we define to be the number of detected galaxies having measured magnitudes and effective radii within 10 per cent of their input values.  And finally, we produce an `error vector' diagram showing the (3$\sigma$-clipped) mean size and direction of the difference between input and recovered values for the detected objects originating in each bin.  Together, the reliability and error vector diagrams allow us to identify any regions of the observable parameter space contaminated by galaxies with measured magnitudes and half light radii that poorly reflect their true, intrinsic properties.  Equivalent plots for our exponential ($n=1$) profile shapes were already presented in \citet{cam07}, although here we have tightened the tolerance on flux and magnitude recovery from 25\% to 10\%.  Thus, we recompiled our previous $n=1$ profile results with this new tolerance so as to be able to indicate the corresponding 75\% completeness and reliability limits for this profile shape for comparison in Fig.\ \ref{dev}.\\

It is clear from our completeness diagram that de Vaucouleurs ($n=4$) profile shape galaxies are easily detectable over the entire area of the apparent magnitude-half light radius plane spanned by the observed iUDF-BRIGHT distribution.  Their highly centrally-concentrated light distributions ensure that even for low mean surface brightness systems there is still a compact, high surface brightness core that emerges above the background sky.  Hence, they are more readily detectable than exponential profile galaxies which we fail to identify in the UDF $i$-band image beyond $\sim$28.0 mag arcsec$^{-2}$.  However, when we investigate how accurately we can measure the intrinsic magnitudes and sizes of these galaxies the situation is reversed.  It is actually extremely difficult to recover the intrinsic flux of a de Vaucouleurs profile system as a significant fraction of the light is contained in its low surface brightness outer regions (the profile `wings').  Whilst we can recover intrinsic magnitudes and sizes of exponential profile galaxies to within 10\% accuracy down to $\left<\mu\right>_{e}$$\sim$27.0 mag arcsec$^{-2}$, here we find we can only recover those of de Vaucouleurs profile galaxies down to $\sim$26.0 mag arcsec$^{-2}$.  So one might wonder, \textit{if} there exists a significant population of high S\'ersic index systems \textit{intrinsically} occupying the region of the apparent magnitude-size plane in between our completeness and reliability limits, where do they turn up in the \textit{observed} distribution?  For this we inspect the `error vector' diagram, which indicates \textit{they are erroneously recovered as faint, compact objects}.  This illustrates the importance of using a surface brightness selection limit based on measurement reliability as well as detection completeness.\\

The reliability plot also reveals constraints on our ability to accurately observe very large, very small and very faint galaxies.  The properties of galaxies larger than $\sim$$4''$ in the UDF are difficult to measure as they overlap with numerous other sources, causing problems with deblending.  Galaxies smaller than $\sim$$0''.06$ pose a challenge for accurate measurements of the half light radius as they are a similar size to the PSF ($\sim$$0''.081$) and pixel scale ($0''.03$).  Also, we cannot measure reliable magnitudes for objects fainter than 28.0 mag in the $i$-band image.  This is because, even though 28.0 mag is well above the 10$\sigma$ point source detection limit of 29.2 mag, the signal-to-noise ratio is low outside the cores of faint, extended sources.  Due to the decision to locate the UDF away from known, nearby, bright objects there is also an effective selection limit on the brightest galaxies in the sample.  One can estimate this limit from the brightest apparent (galaxy) magnitude in the catalogue, which is 18.26 mag in the $i$-band.\\

By considering all the limits described above we define an observational window in the apparent magnitude-size plane inside which all galaxies are detected with high completeness and structural parameters are reliably recovered.  We shall refer to this as the `bias-free' region.  This five-sided, bias-free region is indicated in Fig.\ \ref{dev}.\\

\subsection{Absolute Selection Limits}
The selection boundaries derived for the apparent luminosity-size plane are easily extended to the absolute regime for a volume-limited sample via the method of \citet{dri99}.  The relevant equations are presented and explained in \citet{cam07}, and include the effects of both the $(1+z)^4$ cosmological surface brightness dimming and the K-correction.  In this study we use a series of 5 redshift bins for our volume-limited samples spanning the redshift range $z=0.25$ to 1.5.  Our absolute selection limits for galaxies in each redshift bin are marked on the luminosity-size diagrams of Fig.\ \ref{mag_re}.  Note that we apply the $n=1$ limits to our `blue, diffuse' class and the $n=4$ limits to our `red, compact' class.\\

\begin{figure*}
\center
\includegraphics[width=18.5cm,height=5.7cm]{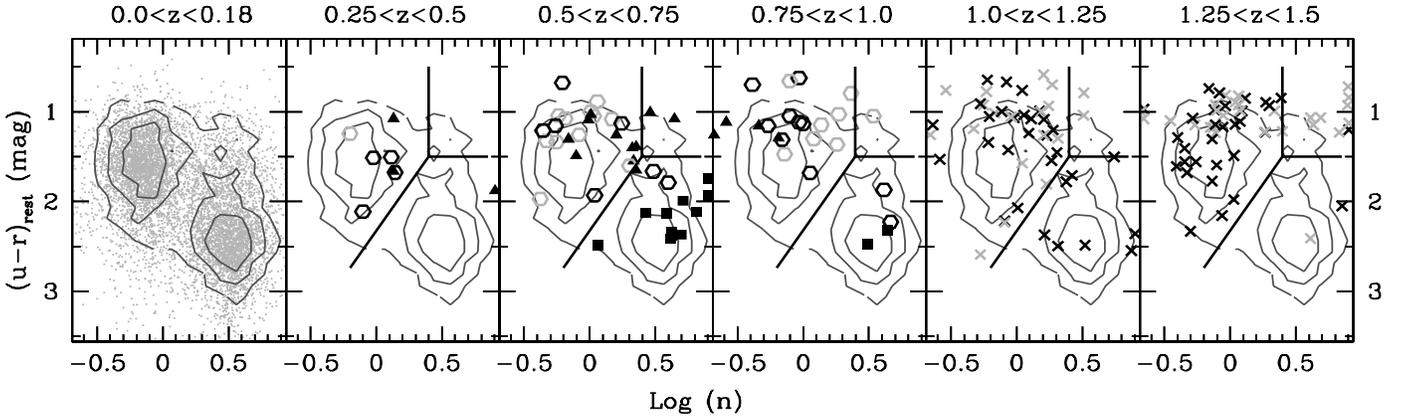}
\caption{\label{color_logn}Color-log($n$) diagrams for the MGC $0.0 < z < 0.25$ sample and the iUDF-BRIGHT volume-limited samples to $z = 1.5$.  The MGC local galaxy distribution is clearly bimodal (as demonstrated by \citealt{dri06}) and its number density contours are overlaid on the high redshift UDF panels for comparison as a local benchmark.  The UDF distributions are not evidently bimodal, except perhaps at $0.75 < z < 1.0$, for reasons discussed in Section \ref{errors}, including cosmic variance and measurement errors.  Eyeball morphological classifications for iUDF-BRIGHT galaxies at $z < 1.0$ are indicated with symbols corresponding to \textit{solid square} = E/S0, \textit{solid triangle} $=$ Sa-Sc and \textit{open hexagon} $=$ Sd/Irr/Merger.  The solid lines indicate our division of the sample into three categories, namely `blue diffuse', `red compact' and `blue compact'.  Objects in all panels are subject to a constant faint magnitude limit of $M_B < -18$ mag.  We also explore the effect of brightening this limit to account for the passive ageing of galaxy stellar popuations (see Section \ref{errors}).  Objects fainter than the evolving magnitude limit are colored grey instead of black.}
\end{figure*}

\section{Results}
\subsection{Color-Log($n$) Diagrams}
In Fig.\ \ref{color_logn} we present the iUDF-BRIGHT color-log($n$) distribution of bright galaxies ($M_B < -18$ mag) in five volume-limited samples from $z=0.25$ to 1.5.  We also plot the color-log($n$) distribution of the 7877 bright galaxies ($M_B < -18$ mag) at $0.0 < z < 0.25$ in the Millennium Galaxy Catalogue (MGC) \citep{lis03,dri05} for comparison as a low redshift benchmark.  \citet{dri06} demonstrate that the MGC local galaxy sample shows clear bimodalities in both rest-frame $(u-r)$ color and global S\'ersic index, and that there exist two extremely well separated populations in the color-log($n$) plane.  This bimodality in the benchmark MGC distribution is highlighted in Fig.\ \ref{color_logn} by volume-density contours overlaid against the raw MGC distribution in the first panel and against the iUDF-BRIGHT high redshift samples in the subsequent panels.  The saddle point between the two peaks offers an objective, physically-motivated subdivision of the galaxy population.  We hoped to detect the equivalent bimodality in our high redshift bins, thereby ensuring self-consistent classification over all epochs.  However, even with the MGC contours guiding the eye, no clear bimodality is evident in the iUDF-BRIGHT samples, except perhaps at $0.75 < z < 1.0$.  This is presumably due to the combined effects of cosmic variance, luminosity evolution, measurement uncertainties and the morphological K-correction as we discuss below.  The final divisions we use to separate our sample into `blue, diffuse', `red, compact' and `blue, compact' classes are indicated by thick black lines in Fig.\ \ref{color_logn}.  The choice of these divisions is explained in Section \ref{subdiv}\\

\subsection{Sources of Uncertainty and Bias}\label{errors}
It is important to bear in mind that the results presented here are highly susceptible to cosmic variance (see \citealt{som04}). Given the UDF field of view of 11 arcmin$^{2}$ we note that the volumes surveyed within the various redshift intervals shown in Fig.\ 2 are: 608 Mpc$^{3}$ ($0.25 < z < 0.50$); 1261 Mpc$^{3}$ ($0.50 < z < 0.75$); 1860 Mpc$^{3}$ ($0.75 < z < 1.00$); 2332 Mpc$^{3}$ ($1.00 < z < 0.25$); and 2674 Mpc$^{3}$ ($1.25 < z < 1.50$). Within such small volumes cosmic variance can be severe (up to 200\%, c.f. \citealt{som04}, Fig.\ 2) and any trends could therefore be due to both evolutionary {\it and} environmental effects.  Based on random sampling of the local MGC survey one typically requires a volume of order 100,000 Mpc$^{3}$ to reduce cosmic variance to $\sim$10\%, i.e., $\sim$50 UDFs.\\

One of the weaknesses of the color-log($n$) bimodality as a tool for identifying galaxy types is its sensitivity to the magnitude limit imposed.  The original $M_B < -18$ mag cut was chosen as it gives roughly equal numbers of red and blue peak galaxies in the local MGC sample.  With a brighter limit the red galaxies dominate the sample and with a fainter limit the blue galaxies dominate.  Numerous studies (recently \citealt{bar05,dah05,ilb06,mci05,zuc06}) have found evidence for a decrease in the characteristic luminosities of different galaxy types from $z \sim 1.5$ to the present, attributed primarily to the ageing of their stellar populations.  Hence, our fixed absolute magnitude limit of $M_B < -18$ mag may include some galaxies at high redshift that would become too faint over time to make the cut at low redshift.  We attempt to account for this effect by correcting our high redshift limits according to the anticipated luminosity evolution (as in \citealt{bar98}).  Specifically, we adopt corrections of $\Delta M = -0.5 z$ for our `red, compact' class limit, $\Delta M = -0.6 z$ for our `blue, diffuse' class limit and $\Delta M = -1.75 z$ for our `blue, compact' class limit, movitated by the results of \citet{zuc06} and \citet{ilb06}.  Any points excluded by these additional constraints are colored grey (instead of black) in Fig.\ \ref{color_logn} and Fig.\ \ref{mag_re}.  A large number of faint, blue galaxies are removed by this cut, but no significant increase in bimodality is evident.  Interestingly, all but one of our genuine `blue, compact' objects would likely be fainter than the $M_B < -18.0$ mag cut locally.  It is likely these galaxies are the pre-cursors of modern day intermediate and dwarf galaxies caught at an epoch of intense star-formation.\\

Measurement errors in individual S\'ersic indices and colors contribute only a small source of uncertainty to the recovered UDF galaxy distributions.  \citet{coe06} estimate an error on the optimum model fit to each galaxy using \textsc{galfit}, which for our sample has a median value of $\widetilde{\Delta n} \sim 0.2$.  They also calculate uncertainties on their aperture magnitudes based on image noise, and the corresponding median values for our sample are $\widetilde{\Delta B} = 0.01$, $\widetilde{\Delta V} = 0.06$, $\widetilde{\Delta i} = 0.04$ and $\widetilde{\Delta z} = 0.05$ mag.  This implies median errors on our apparent $(B-i)$ and $(V-z)$ colors of 0.05 and 0.08 mag respectively.  Additional errors of $\la$0.04 mag are expected from our K-corrections given the SED template and band-pass shifting approach adopted.  The total uncertainties on our rest-frame $(u-r)$ colors are thus $\la$0.09 mag.\\

Finally, we note the problem of the morphological K-correction, i.e., the intrinsic variation in galaxy morphology with rest-frame wavelength.  \citet{coe06}'s S\'ersic fits are all performed in the observed $i$-band, which at $z \sim 0.75$ samples close to rest-frame $B$-band (optical) light, but by $z \sim 1.5$ samples closer to rest-frame $U$-band (ultra-violet) light.  The dominance of UV emission by bright, young stars is known to cause galaxy images to appear clumpier and of later type than they do at optical wavelengths \citep{kuc00,tay07} and galaxy light profiles to appear less concentrated.  For instance, \citet{lab03} find mean variations and standard deviations in S\'ersic index from optical to UV of $\left<\Delta \log(n)\right> = -0.09$, $\sigma = 0.10$ and $\left<\Delta \log(n)\right> = -0.02$, $\sigma = 0.19$ for samples of discs and spheroids respectively at $z \sim 0.21$.  This effect may contribute towards obscuring any intrinsic bimodality in the color-log($n$) distributions of UDF galaxies in the lowest and highest redshift bins of Fig.\ \ref{color_logn}.\\

\subsubsection{Bimodality at $0.75 < z < 1.0$}
Given the potential sources of uncertainty and bias that may distort our measured galaxy color-log($n$) distributions at high redshift, it is perhaps no surprise that most of the UDF volume-limited samples fail to reveal strong evidence of bimodality.  However, in the $0.75 < z < 1.0$ bin where the morphological K-correction is negligible there does appear to exist two distinct (well separated) populations---one displaying the blue colors of star-forming galaxies and the diffuse light profiles (low S\'ersic indices) of disc-dominated systems, and the other displaying the red colors of old stellar populations and the compact light profiles (high S\'ersic indices) of bulge-dominated systems.  However, as mentioned earlier, the small volume sampled by the UDF over this redshift interval leaves this result highly susceptible to cosmic variance.  Further investigation of this issue will thus be warranted once a suitably large, deep imaging dataset becomes publicly available.\\

\subsubsection{Sample Division and Morphological Correlation}\label{subdiv}
\begin{figure*}
\center
\includegraphics[width=14cm]{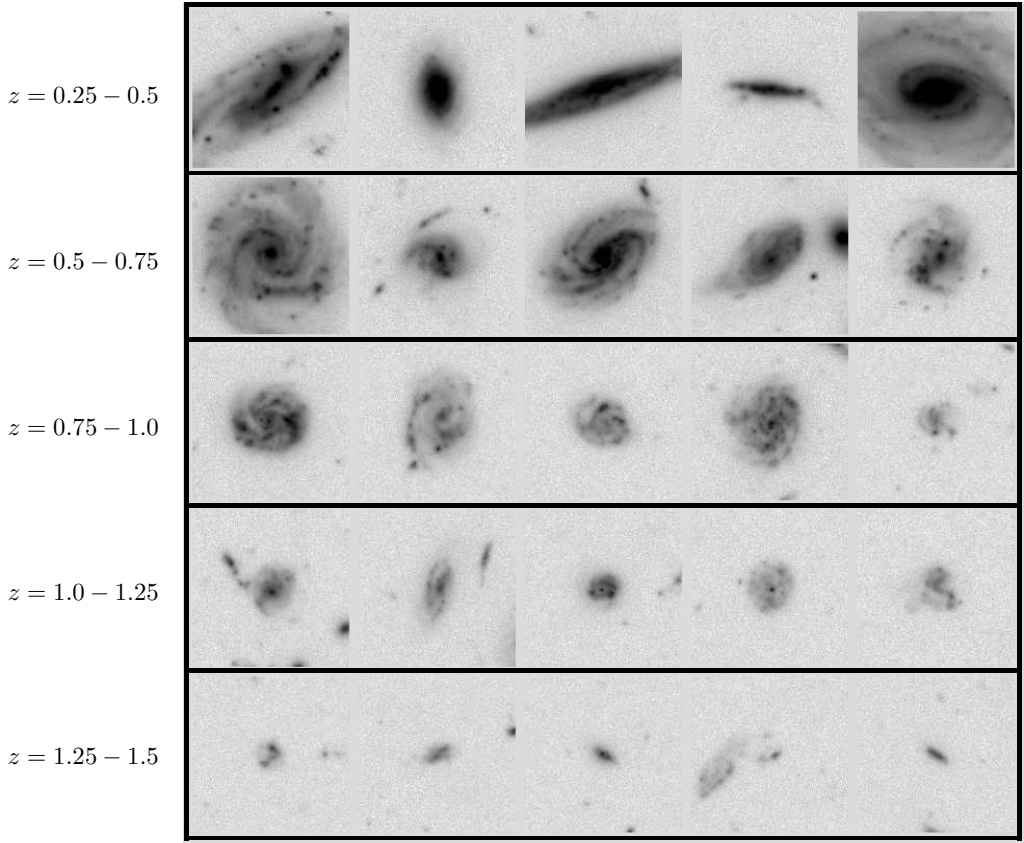}\\
\caption{\label{blue_diffuse}Galaxy morphology in our `blue, diffuse' class.  Each row contains postage stamp images showing members of this class cut from the ACS/WFC $i$-band UDF image.  These are clearly disc-dominated systems, predominantly late-type spirals and irregulars.  The rows are ordered in increasing redshift intervals from $0.25 < z < 0.50$ to $1.25 < z < 1.5$ to illustrate the changes in galaxy appearance due to bandpass shifting, decreasing angular sizes and morphological evolution.}
\end{figure*}

\begin{figure*}
\center
\includegraphics[width=14cm]{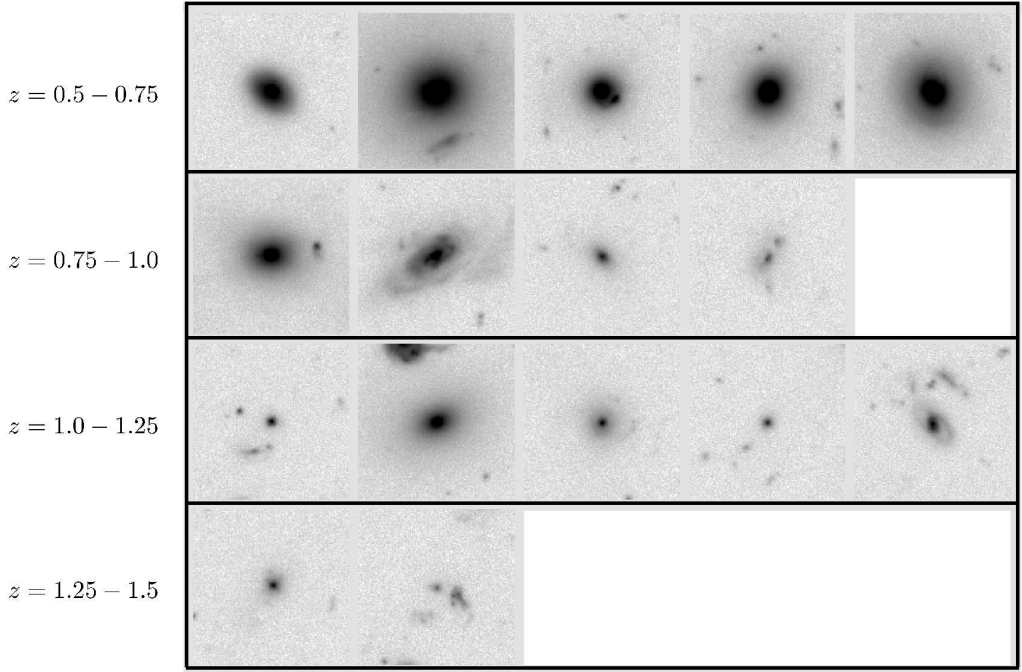}\\
\caption{\label{red_compact}Galaxy morphology in our `red, compact' class.  Each row contains postage stamp images showing members of this class cut from the ACS/WFC $i$-band UDF image.  These are clearly bulge-dominated systems, mainly giant ellipticals and S0/Sa galaxies.  The rows are ordered in increasing redshift intervals from $0.5 < z < 0.75$ to $1.25 < z < 1.5$ to illustrate the changes in galaxy appearance due to bandpass shifting, decreasing angular sizes and morphological evolution.}
\end{figure*}

\begin{figure*}
\center
\includegraphics[width=14cm]{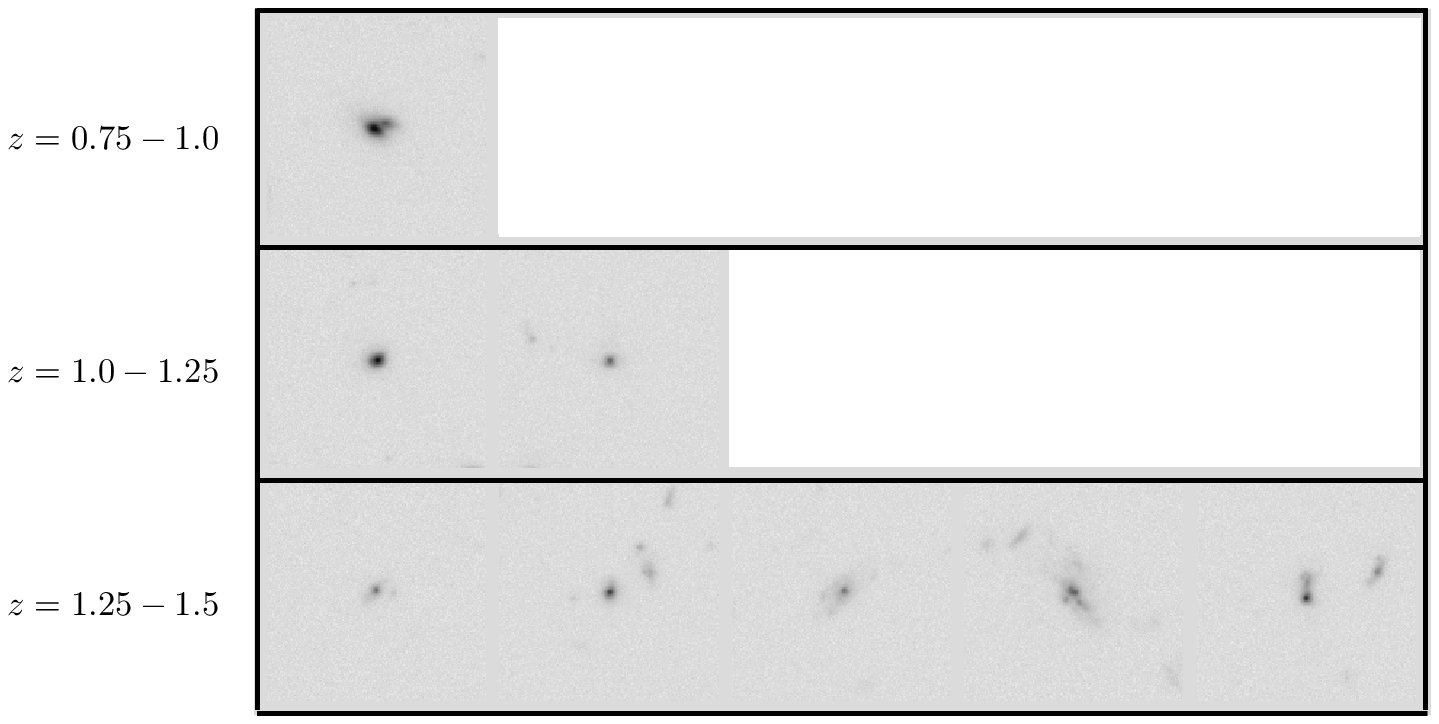}\\
\caption{\label{blue_compact}Galaxy morphology in our `blue, compact' class.  Each row contains postage stamp images showing members of this class cut from the ACS/WFC $i$-band UDF image.  These are all small nucleated systems, featuring surrounding regions of irregular, diffuse material and only appear in our sample at high redshift.  The rows are ordered in increasing redshift intervals from $0.75 < z < 1.0$ to $1.25 < z < 1.5$ to illustrate the changes in galaxy appearance due to bandpass shifting, decreasing angular sizes and morphological evolution.}
\end{figure*}

Our main aim in constructing the UDF high redshift color-log($n$) diagrams was to evaluate their suitability as a tool for subdividing the galaxy population.  Unfortunately, the lack of clear evidence of bimodality in these high redshift distributions means that our classification approach cannot be defined self-consistently from the galaxy populations at each epoch.  Hence, we adopt the saddle point of the local bimodality, $(u-r)_{\mathrm{rest}} = 2.325 - 2.074 \log(n)$, to separate the two key galaxy types, i.e., our `blue, diffuse' and `red, compact' classes.  Visual inspection of the data reveals that this is still an appropriate division of the two distinct populations apparent in the $0.75 < z < 1.0$ bin.  We note also note that the steepness of this line in color-log($n$) space reduces the sensitivity of our cut to the observed evolution in galaxy colors.  Postage stamp images showing examples of each of these galaxy classes are shown in Fig.\ \ref{blue_diffuse} and Fig.\ \ref{red_compact}.  In addition we classify galaxies in the region described by $\log(n) > 0.398$ and $(u-r)_{\mathrm{rest}} < 1.5$ as `blue, compact' systems.  These systems are common in our sample at high redshift, but are unmatched by bright local counterparts in the MGC.  Examples of this type of galaxy are shown in Fig.\ \ref{blue_compact}.  They all exhibit a bright, compact nucleus embedded in an extended region of diffuse, irregular emission.  Such galaxies have been observed in other surveys, predominantly at high redshift (e.g.\ \citealt{im01,men04,cro04,ilb06,ell05,dri07}), but also locally (e.g.\ \citealt{dri05}).  When the evolving luminosity limit was introduced, all but one of these galaxies were too faint for the cut.  These galaxies are presumably not the direct pre-cursors of modern day giant systems, but are probably lower mass systems caught at an epoch of intense star-formation.\\

Although we are attempting to implement an objective, quantitative method of classification, it is still informative to compare our cuts against the distribution of visual morphological types.  \citet{dri06} reported a strong correlation between visual morphology and color-log($n$) plot position.  Namely, that early-types occupy the red, compact peak and late-type spirals and irregulars occupy the blue, diffuse peak with intermediate spirals bridging the gap.  Hence, we inspected the ACS $B$, $V$, $i$ and $z$-band postage stamp images of all UDF galaxies in our $M_B < - 18$ mag sample with $z < 1$ and assigned them to one of three groups based on their Hubble type: E/S0, Sa-Sc, or Sd/Irr.  Different symbols are used for each galaxy in our color-log($n$) diagrams of Fig.\ \ref{color_logn} to indicate these types.  Our high redshift galaxies echo the findings of \citet{dri06} regarding the relation between color-log(n) position and visual morphology at low redshift.  Furthermore, we see that our chosen division line between `blue, compact' and `red, compact' types is indeed effective at separating E/S0s from Sa-Sc and Sd/Irrs.  We do not attempt to assign eyeball morphologies to the $1.0 < z < 1.5$ objects as these are more difficult to classify, being less well resolved in our images and dominated by irregular features.  This is due both to intrinsic evolution in the star-formation rate and the fact that the $i$ and $z$-bands are sampling rest-frame UV light (i.e., young stellar populations) at these redshifts.  Even so, when we examine $z > 1$ galaxies from each peak alongside their counterparts at $z < 1$ there is a clear consistency in appearance, as seen in Fig.\ \ref{blue_diffuse} and Fig.\ \ref{red_compact}.  Overall, this investigation of the visual morphologies of galaxies in our sample gives us confidence in our method of subdivision on the color-log($n$) plane.\\

\subsection{Quantifying Color Evolution}
A certain degree of color evolution is clearly evident in the color-log($n$) distributions of Fig.\ \ref{color_logn}.  Specifically, the mean color of our `blue, diffuse' galaxies becomes substantially redder from $\left< (u-r)_{\mathrm{rest}}\right> = 1.12 \pm 0.09$ mag in the UDF at $0.75 < z < 1.0$ and $1.28 \pm 0.09$ mag in the UDF at $0.5 < z < 0.75$ to $1.59 \pm 0.01$ mag in the MGC at $0 < z < 0.25$.  This represents a change of $0.62 \pm 0.10$ mag from $z=1$ to $z=0$.  The `red, compact' galaxies also become redder on average, but not as strongly, evoluting from $\left< (u-r)_{\mathrm{rest}}\right> = 2.21 \pm 0.09$ mag in the UDF at $0.75 < z < 1.0$ and $2.08 \pm 0.09$ mag in the UDF at $0.5 < z < 0.75$ to $2.42 \pm 0.01$ mag in the MGC at $0 < z < 0.25$.  This represents a change of $0.37 \pm 0.10$ mag from $z = 1$ to $z = 0$.  These results are in line with those of \citet{bla06} comparing objects at $z \sim 1$ in the DEEP2 survey to those found in the nearby universe in the SDSS.  In that study blue sequence galaxies were found to be redder in ${}^{0.1}$$(g-r)$ color (i.e., $(g-r)$ color using a rest-frame of $z=0.1$) by 0.3 mag by $z \sim 0$, while red sequence galaxies were redder by 0.1 mag.  In ${}^{0.1}$$(g-r)$ color the redshifted 4000$\AA$ break lies within the $g$-band whereas in $(u-r)_{\mathrm{rest}}$ the two filters lie either side of it.  This means that $(u-r)_{\mathrm{rest}}$ color is more sensitive to changes in the size of the 4000$\AA$ break.  Using our SED template fits and filter throughput functions we can compute ${}^{0.1}$$(g-r)$ colors for our galaxies.  In this way we find our $(u-r)_{\mathrm{rest}}$ color changes are equivalent to ${}^{0.1}$$(g-r)$ color changes of $0.31 \pm 0.10$ mag and $0.21 \pm 0.10$ mag for our `blue, diffuse' and `red, compact' populations respectively.\\

\subsection{The Luminosity-Size Relation by Galaxy Type}\label{evol}
\begin{figure*}
\center
\includegraphics[width=18.5cm,height=9.0cm]{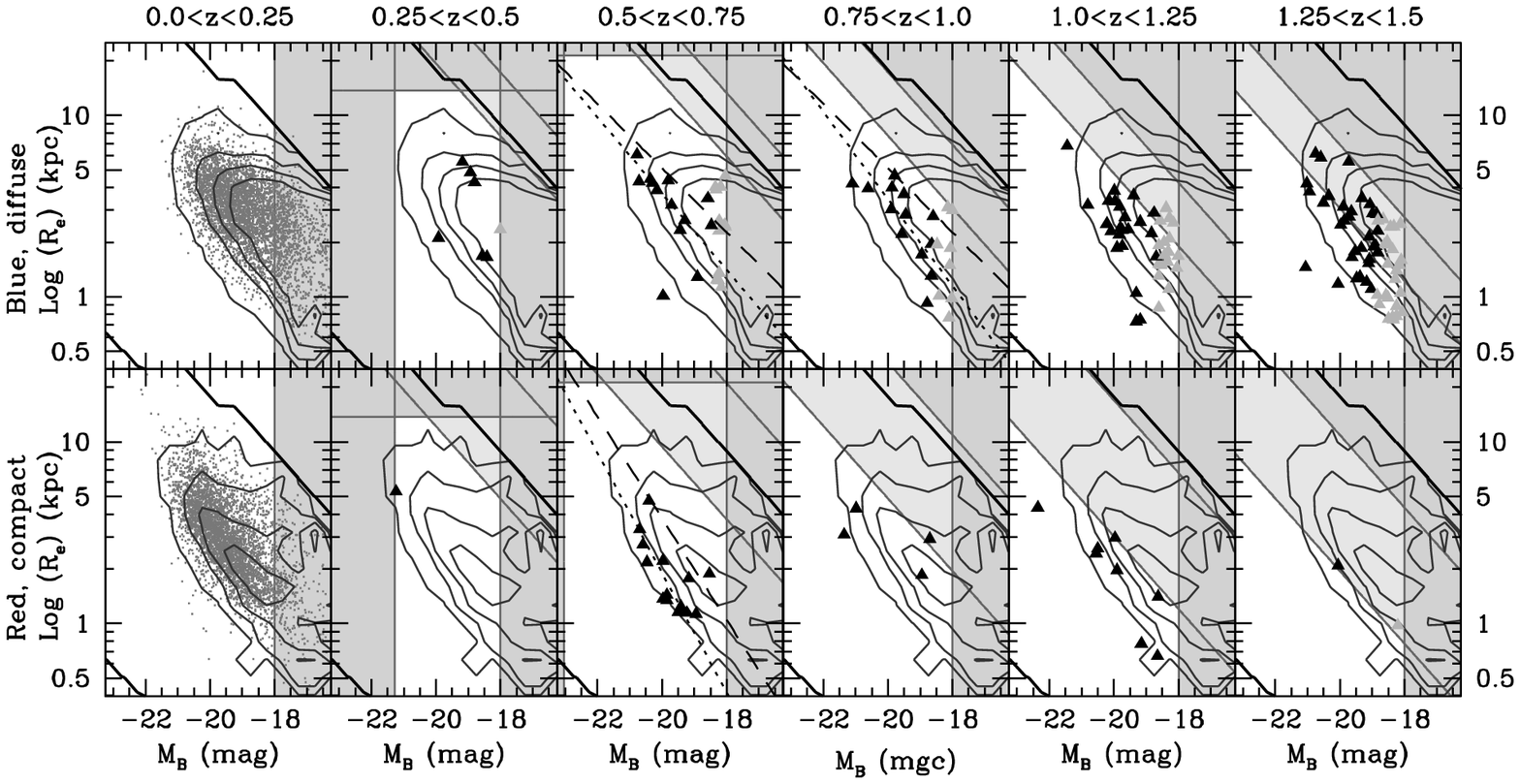}
\caption{\label{mag_re} Luminosity-size relations by galaxy type using our `blue, diffuse' (\textit{top row}) and `red, compact' (\textit{bottom row}) classes for the MGC $0.0 < z < 0.25$ sample and the iUDF-BRIGHT volume-limited samples to $z = 1.5$.  The two classes are seen to occupy distinctly different regions of this parameter space in the nearby universe (MGC sample).  The `red, compact' luminosity-size relation shows a higher mean surface brightness and a steeper slope than that for the `blue, diffuse' type.  These MGC relations are fit with contours and overlaid in grey on the high redshift iUDF-BRIGHT samples.  Selection limits on object detection and reliability for the UDF survey are indicated with dark lines and grey shading.  In the remaining whitespace areas galaxies are able to be detected with $>$90\% completeness, and their sizes and magnitudes are able to be measured to within 10\% of their true Kron values for more the 75\% of cases.  These boundaries were computed as described in Section 3, using the limit appropriate to $n=1$ profile shapes for the diffuse types and that for $n=4$ shapes for our compact types.  The MGC selection boundary corresponding to an iso-volume contour at 100 Mpc$^3$ is indicated with a thick black line.  The bias-free regions inside the relevant selection limits of the two surveys are comparable for our volume-limited samples out to $z\sim 1$, at which point the observed distributions become overwhelmed by surface brightness-dependent biases.  At $z < 1$ though, the UDF galaxies appear to show evidence for evolution towards higher mean surface brightnesses and we investigate this in Section \ref{evol}.}
\end{figure*}

In Fig.\ \ref{mag_re} we present the $B$-band, iUDF-BRIGHT luminosity-size relation by galaxy type, i.e., for our `red, compact' and `blue, diffuse' classes, in our five volume-limited samples to $z=1.5$.  The relevant selection limits on completeness and measurement reliability for high and low $n$ systems (as calculated in Section 3) are marked with grey shading on each plot.  Objects removed by our evolving magnitude limits are again colored grey instead of black.  The equivalent MGC local galaxy relations were constructed by adopting the same color-log($n$) criteria as for the UDF samples.  An (inverse) observable volume-weighted correction was then applied to generate the MGC number density contours.  These are overlaid on all the high redshift panels for comparison as a local benchmark.  The 100 Mpc$^3$ iso-volume contour, within which MGC selection biases are well constrained, is indicated with a thick black line.\\

It is clear from these diagrams that the UDF survey's bias-free observational window covers a similar region of parameter space out to redshift unity as that which the MGC does locally.  With the exception of the $0.75 < z < 1.0$ `red, compact' bin, all our UDF luminosity-size distributions at $z < 1$ are well separated from the relevant surface brightness selection limits.  Hence, we are confident that the sizes and magnitudes measured for these objects are free of surface brightness-dependent biases within a 10\% tolerance.  At these redshift intervals the `blue, diffuse' UDF galaxies appear to show evidence of evolution towards higher $B$-band surface brightnesses relative to the $z \sim 0$ MGC population.  Quantitatively, we measure mean $\left< \mu \right>_{e}$ of $21.3 \pm 0.2$ and $21.1 \pm 0.2$ mag arcsec$^{-2}$ for galaxies brighter than the evolving magnitude limit in the UDF $0.5 < z < 0.75$ and $0.75 < z < 1.0$ bins respectively.  This is compared to $22.20 \pm 0.01$ mag arcsec$^{-2}$ in the MGC.  (The error margins are derived from the sum in quadrature of the uncertainties on the apparent magnitudes, sizes and redshifts for an individual UDF galaxy).  This is a $1.57 \pm 0.22$ mag arcsec$^{-2}$ increase from $z=0$ to $z=1$ (based on the line of best fit between the one MGC and two high redshift UDF datapoints) and is in good agreement with recent results from other high redshift studies.  For instance, \citet{bar05} find evolution of $1.43 \pm 0.07$ mag arcsec$^{-2}$ to $z = 1$ in rest-frame $B$-band for an $M_V < -20$ mag disc galaxy sample in the GEMS survey data.  The slightly smaller amount of evolution they find could be due to the brighter limit of their sample as there is some evidence for a luminosity (or size) dependence to the surface brightness evolution in our data.  Least squares fits to the luminosity-size relations in the UDF $0.5<z<0.75$ and $0.75<z<1.0$ bins return gradients of $-0.21 \pm 0.06$ and $-0.24 \pm 0.06$ log(kpc)/mag respectively, compared to $-0.177 \pm 0.004$ log(kpc)/mag for the MGC local relation.  Other studies have yielded similar results.  For instance, \citet{mel07} report $B$-band luminosity evolution of $1.53 \pm^{0.13}_{0.10}$ mag to $z=1$ for large ($R_e > 5$ kpc) blue galaxies in GOODS-N and $1.65 \pm^{0.08}_{0.18}$ mag for intermediate ($3 < R_e < 5$ kpc) ones.  Beyond $z \sim 1$ our surface brightness-dependent limits on UDF object detection and parameter reliability impinge on the observed distribution and the valid region of comparison to the MGC data.  Thus, although the mean $\left< \mu \right>_{e}$ continues to increase to $\sim$20.7 mag arcsec$^{-2}$ by our $1.0<z<1.25$ bin, we cannot distinguish between real evolution and the impact of selection effects at these redshifts.\\

The high redshift luminosity-size relations for our `red, compact' sample are somewhat more difficult to interpret.  Even though the selection limits are significantly tighter on these higher S\'ersic index systems (as shown in Section 3.1), our UDF $0.5 < z < 0.75$ bin still samples an equivalent region of parameter space to the MGC survey, and our galaxies are detected and measured free of bias.  One can see from Fig.\ \ref{mag_re} that the luminosity-size relation of the local MGC `red, compact' population has a higher surface brightness and steeper slope than the `blue, diffuse' one, which partially compensates for the tighter selection limits.  This is also the case in our UDF $0.5<z<0.75$ bin where the mean `red, compact' $\left< \mu \right>_{e}$ is $20.1 \pm 0.2$ mag arcsec$^{-2}$, i.e., 1.2 mag arcsec$^{-2}$ brighter than the `blue, diffuse' galaxies at this redshift.  It is also an increase in surface brightness compared to the MGC local sample of `red, compact' systems, which has a mean $\left< \mu \right>_{e}$ of $21.65 \pm 0.01$ mag arcsec$^{-2}$.  Extrapolating this trend with a linear fit gives an evolution of $\sim$3 mag arcsec$^{-2}$ to $z=1$, which is much faster evolution than would be expected for a population usually found to have observational properties consistent with passively fading, old stellar populations.  However, we cannot confirm this unusual result with our $0.75<z<1.0$ bin as we have just four galaxies in our sample and their mean surface brightness of $20.9 \pm 0.2$ mag arcsec$^{-2}$ suggests weaker evolution of only $1.0 \pm 0.4$ mag arcsec$^{-2}$ to $z=1$.  A line of best fit anchored to the MGC value and passing between these two conflicting data points gives an evolution of $1.65 \pm 0.22$ mag arcsec$^{-2}$ from $z=0$ to $z=1$.\\  

When moderated by inclusion of the $0.75 < z < 1.0$ data as above our results are in good agreement with those of previous studies.  For instance, \citet{mci05} examined the luminosity-size relation of red sequence galaxies in GEMS.  They found size-dependent evolution to $z=1$ in $V$-band magnitude of $1.6 \pm 0.1$ mag for small systems ($R_e < 1$ kpc), $1.3 \pm 0.1$ mag for medium systems ($1 < R_e < 2$ kpc) and $0.7 \pm 0.1$ for large systems ($R_e > 2$ kpc) (taking $h=1$ for consistency with our work).  In our $0.5<z<0.75$ bin where we have enough galaxies to fit the luminosity-size relation slope we too find a suggestion of size-dependent evolution with a change in slope to $-0.33 \pm 0.05$ log(kpc)/mag from $-0.281 \pm 0.003$ log(kpc)/mag at $0 < z < 0.25$, but this measurement is also consistent with null evolution within the uncertainties.  The conflict between the exceptionally strong surface brightness evolution detected in our $0.5< z <0.75$ bin and the milder result of \citet{mci05} (or indeed that of our $0.75 < z < 1.0$ bin) is presumably due to field to field differences (i.e., cosmic variance) across our small survey volumes.  Catastrophic errors in the photometric redshifts computed for some of these systems may also be partially responsible.  Finally, the `red, compact' class suffers even more from the impact of selection effects at high redshift than does the `blue, diffuse' class, so again we can gain little insight from our $z > 1$ samples.\\

\section{Discussion}
The bimodal nature of the bright galaxy population in various physical parameters is displayed locally in color-log($n$) \citep{dri06}, log($n$)-magnitude \citep{dri06,bal06} and color-magnitude \citep{str01} space, and at $z \sim 1$ in color-magnitude space \citep{bel04,mci05,bla06}.  This indicates the existence of two distinct galaxy sub-populations: (1) those of our `red, compact' class that exhibit the red colors of old, passively-evolving stellar populations and the compact light profiles typical of spheroidal, velocity dispersion-supported structures, and (2) those of our `blue, diffuse' class that exhibit blue colors indicative of recent star-formation and the diffuse light profiles typical of rotationally-supported systems.  In general, as morphological comparisons indicate (\citealt{dri06} and here), these types may be understood to reflect the two sides of the Hubble sequence, namely ellipticals/lenticulars and spirals/irregulars.  \citet{dri06} have argued that the bimodality specifically reflects the two component nature of galaxies with our two types corresponding to bulge-dominated and disc-dominated systems respectively.  In any case, the existence of the galaxy bimodality suggests that there are two distinct pathways of formation, one leading to `red, compact' types and one leading to `blue, diffuse' types.  Here we discuss these pathways and attempt to account for our results in light of contemporary galaxy formation theory and observation.\\

Hierarchical clustering (HC) theory is the predominant galaxy formation scenario at present as it can link the evolution of the baryonic matter in galaxies to the evolution of their host dark matter haloes \citep{whi78} inside the framework of the $\Lambda$CDM cosmological model \citep{spe07}.  In HC the two pathways of galaxy formation are explained in terms of the evolution of disc galaxies and ellipticals.  Discs are thought to form from the collapse and subsequent cooling of primordial gas clouds inside rotating dark matter haloes (e.g. \citealt{fal80}), spun up by mergers and the tidal torques of their neighbors in the early universe \citep{pee69}.  \citet{gov07} find that in their $\Lambda$CDM-based, semi-analytical models a stellar disc forms in Milky Way-sized halos immediately after their last major merger, which is $z > 2$ (as supported by the dark matter assembly history studies of \citealt{li05}).  At low redshifts disc evolution is characterized by accretion of small satellites, external gas accretion and secular processes such as bar formation.\\

Elliptical galaxies are built up over cosmic time by mergers between similar-sized discs \citep{too72} and perhaps also so-called `dry-mergers' amongst the elliptical population \citep{van99,naa03}. Semi-analytical modeling confirms that the formation history of ellipticals in a $\Lambda$CDM universe follows that of dark matter halos \citep{del06} with the most massive ellipticals and halos assembled at relatively low redshift.  However, \citet{del06} find the bulk of stars in these systems were formed at high redshift in smaller galaxies, which were subsequently incorporated into the primary galaxy by merging.  For instance, for ellipticals of mass $>$$10^{11}$ M$_{\odot}$ ($M_B \sim -20$ mag) the median redshift when half of the stars are formed is $\sim$2.5, yet half of their stars are typically assembled in a single object only at redshift $\sim$0.8.  Hence, the galaxies in our sample ($M_B < -18$ mag) would be expected to show bimodality at $z \sim 1$ in the HC scenario as both types are in place at this redshift, although the bright elliptical population should continue to be built up via merging to low redshift.  This picture is generally supported by observational studies such as \citet{bel04} and \citet{bla06} who find that the color-magnitude bimodality is in place by $z \sim 1$, but that the red sequence continues to evolve significantly in number density to the present day.  In this study we have chosen not to examine evolution by type in the number density or relative fraction of galaxies in different classes as such measurements are especially prone to cosmic variance and selection biases.  However, we can explore the pathways of galaxy evolution in terms of hierarchical clustering theory using the surface brightness and color evolution we have quantified for each type.\\

For our `blue, diffuse' (i.e., disc-like) systems we measured a reddening in $(u-r)_{\mathrm{rest}}$ color of $0.62 \pm 0.10$ mag and a decrease in surface brightness from $z=1$ to $z=0$ of $1.57 \pm 0.22$ mag arcsec$^{-2}$ in the $B$-band.  The star-formation history of disc galaxies in the HC scenario is governed by the continual accretion of gas and dark matter into the parent halo, peaking at high redshift but continuing on to the present day.  \citet{wes02} have used a HC motivated model to predict $(B-V)_{\mathrm{rest}}$ color changes from $\sim$0.30 mag at $z=1$ to $\sim$0.65 mag at $z=0$ for a disc at a 60 degree inclination angle.  Using our SED template fits we convert our measured $(u-r)_{\mathrm{rest}}$ colors to $(B-V)_{\mathrm{rest}}$ and find evolution from $0.36 \pm 0.10$ at $z=1$ to $0.59 \pm 0.22$ mag at $z=0$ in good agreement with their results.  We can also use this color evolution to compute a change in mass to light ratio (M/L), and thereby roughly disentangle luminosity changes from size changes in the mean surface brightness evolution.  Using \citet{bel03}'s Table 7 we find our measured color change corresponds to a change in log(M/L) of $0.41 \pm 0.11$ in the $B$-band, or a decrease in luminosity at fixed mass of $1.03 \pm 0.28$ mag.  This implies that an increase in mean disc size of $22 \pm 14$\% from $z=1$ to $z=0$ is required to account for the $1.57 \pm 0.22$ mag arcsec$^{-2}$ surface brightness evolution we observe.  \citet{som06} have simulated the evolution of disc scale-lengths in a CDM universe.  They find that although the host dark matter halos of discs continue to grow past $z \sim 1$ the build up of halos from the inside-out means that the inner part of the halo where the baryons are concentrated changes little over that time.  Their final prediction of 15-20\% size evolution agrees well with our result.\\

For our `red, compact' type we measured a reddening in $(u-r)_{\mathrm{rest}}$ color of $0.37 \pm 0.10$ mag and a decrease in surface brightness from $z=1$ to $z=0$ of $1.65 \pm 0.22$ mag arcsec$^{-2}$ in the $B$-band.  Due to increases in computational power it has recently become possible to follow the evolution of disc and bulge sizes, as well as masses, in HC theory using semi-analytical techniques.  \citet{alm07} adopt an approach in which spheroid sizes are computed following a merger by applying conservation of energy and the virial theorem.  These authors are able to reproduce many features of the local elliptical galaxy population, such as the Fundamental Plane, and offer specific predictions for evolution of the elliptical luminosity-size relation with redshift, which are illustrated in their Fig.\ 17.  Essentially, they find luminous elliptical galaxies are lower in surface brightness by $\sim$1 mag arcsec$^{-2}$ in the $B$-band at the present day compared to $z=1$, which is less than the $1.65 \pm 0.22$ mag arcsec$^{-2}$ we measure here.  They note that their predicted surface brightness evolution is almost entirely due to a change in M/L of their passively-evolving stellar populations by a factor of $\sim$3 over this time.  Using Table 7 of \citet{bel03} again we can convert our measured color evolution to a M/L evolution, finding a change of $\Delta$log(M/L) = $0.26 \pm 0.11$ (i.e., a factor of $\sim$1.8) between $z=1$ and $z=0$, or an decrease in luminosity at fixed mass of $0.65 \pm 0.28$ mag.\\

Recent evidence suggests that dry mergers may be an important component in giant elliptical galaxy evolution since $z \sim 1$ \citep{van99}.  \citet{kho06} investigate the formation of elliptical galaxies in the HC scenario via semi-analytical modelling with a unique method of estimating post-merger sizes, motivated by the finding of \citet{spr05} that in elliptical galaxies the effective radius of stars formed during a merger is $\sim$5.7 times smaller than the radius of stars that existed in the pre-merger systems (i.e., formed quiescently).  \citet{kho06} show that the scatter in their model merger-to-quiescent ratios behaves as a function of mass like \citet{she03}'s measured scatter in local elliptical half light radii from the SDSS.  They go on to predict mass-dependent evolution whereby the most massive galaxies evolve the most rapidly in size with redshift because they undergo more dry-mergers (and thus have higher quiescent star fractions) than do less massive ones.  For galaxies in the mass range $5 \times 10^{11}$ M$_\odot <$ M $< 1 \times 10^{12}$ M$_\odot$ (i.e., $\sim$$-22.2 > M_{B} > -23.0$ mag using M/L values based on our $z=1$ colors) they find half light radii are $\sim$55\% larger now than at $z=1$, while for galaxies in the mass range $1 \times 10^{10}$ M$_\odot <$ M $< 5 \times 10^{11}$ M$_\odot$ ($-18.0 > M_{B} > -22.2$ mag) sizes are between $\sim$15 and $\sim$25\% larger.  After subtracting the surface brightness evolution due to change in the M/L ratio we measured via the color change (i.e., $0.65 \pm 0.28$ mag) our results suggest elliptical size increases by $37 \pm 11$\% since $z=1$.  This is still faster than the predictions, but our results certainly favor a model with significant size and luminosity changes.\\

\section{Conclusions}
We find that the color-log($n$) plane can be a useful tool for subdividing the galaxy population in a physically-motivated manner beyond the nearby universe.  We confirm that the locally-observed correlation between morphological type and color-log($n$) plane position is in place by redshift unity.  After separating both the UDF and MGC galaxy populations according to the saddle point in the (well-defined) local bimodality, we construct luminosity-size distributions by galaxy type in a series of volume-limited samples.  By comparison of these distributions at high and low redshift we measure surface brightness decreases of $1.57 \pm 0.22$ and $1.65 \pm 0.22$ mag arcsec$^{-2}$ from $z=1$ to $z=0$ for our `blue, diffuse' and `red, compact' classes respectively.  These results are found to be in good agreement with predictions from recent galaxy formation simulations within the $\Lambda$CDM, hierarchical clustering paradigm.\\

\begin{acknowledgements}
The research of E.\  Cameron is supported in part by funds from ARC DP0451426.  He also wishes to thank St Andrews University for hosting him as a Visiting Scholar.  The Millennium Galaxy Catalogue consists  of  imaging  data  from  the Isaac   Newton  Telescope  and  spectroscopic   data  from  the  Anglo Australian Telescope, the ANU 2.3m,  the ESO New Technology Telescope, the Telescopio Nazionale Galileo, and the Gemini Telescope. The survey has been supported  through  grants  from  the  Particle  Physics  and Astronomy  Research Council (UK)  and  the Australian Research Council (AUS).  The  data  and  data  products  are  publicly  available  from http://www.eso.org/$\sim$jliske/mgc/ or on request from J.\  Liske or S.P.\  Driver. 
\end{acknowledgements}

\bibliographystyle{aa} 
\bibliography{udf2_aa}

\end{document}